\documentstyle[aps,preprint,tighten]{revtex}

\begin{document}
\preprint{\vbox{\hbox{TRI--PP--97--5}
                \hbox{MKPH--T--97--7}
                \hbox{hep-ph/9702394}}}

\title{Muon capture by a proton in heavy baryon chiral perturbation theory}

\author{Harold W. Fearing and Randy Lewis}
\address{TRIUMF, 4004 Wesbrook Mall, Vancouver, British Columbia, Canada 
         V6T 2A3}
\author{Nader Mobed}
\address{Department of Physics, University of Regina, Regina, Saskatchewan,
         Canada S4S 0A2}
\author{Stefan Scherer}
\address{Institut f\"ur Kernphysik, Johannes Gutenberg-Universit\"at,
         J. J. Becher-Weg 45, D-55099 Mainz, Germany}

\date{February 21, 1997}
\maketitle

\begin{abstract}
The matrix element for muon capture by a proton is calculated to ${\cal
O}(p^3)$ within heavy baryon chiral perturbation theory using the new ${\cal
O}(p^3)$ Lagrangian of Ecker and Moj\v{z}i\v{s}.  External nucleon fields are
renormalized using the appropriate definition of the wave function
renormalization factor $Z_N$. Our expression for $Z_N$ differs somewhat from
that found in existing literature, but is the one which is consistent with the
Lagrangian we use and the one which ensures, within our approach, the
nonrenormalization of the vector coupling as required by the conserved vector
current.  Expressions for the standard muon capture form factors are derived
and compared to experimental data and we determine three of the coefficients of
the Ecker - Moj\v{z}i\v{s} Lagrangian, namely, $b_7, b_{19}$, and $b_{23}$.
\end{abstract}

%\narrowtext

Chiral perturbation theory (CHPT) is an effective theory for QCD that allows 
systematic calculations to be performed whenever
external momenta are small with respect to the chiral symmetry breaking scale, 
$\Lambda_\chi \sim 1$ GeV.
The theory was originally formulated for light mesons only\cite{GL}, but
heavy baryons can also be included without sacrificing the small-momentum
expansion\cite{JM}.
As for any effective theory, the Lagrangian of CHPT contains parameters
which are not determined by CHPT itself, but which can be inferred from 
experimental data.  Recent reviews of the vast amount of work that has been
done with CHPT can be found in Refs.~\cite{Ecker95,BKM95,Pich,Bijnens}.

The complete Lagrangian for a single nucleon coupling to pions and 
external fields up to third 
order in small momenta (denoted ${\cal L}^{EckM}_{\pi N}$) has only recently 
been constructed by Ecker and Moj\v{z}i\v{s}\cite{EM96}, although calculations 
for specific processes had been performed earlier.  
In the present work, we study muon capture by a proton with the new 
Lagrangian, ${\cal L}^{EckM}_{\pi N}$.
The form factors that appear in the muon capture amplitude have been considered
previously within heavy baryon CHPT\cite{BKM95,BKKM}, but not with the new
Lagrangian ${\cal L}^{EckM}_{\pi N}$.

Our calculation gives explicit expressions for each of the muon capture
form factors, in terms of parameters that appear in ${\cal L}^{EckM}_{\pi N}$.
We use experimental data to determine the numerical values of the parameters,
which are directly transferable to future calculations
of other processes where ${\cal L}^{EckM}_{\pi N}$ is used.  In particular, 
the parameters of the present work are a subset of the ones that appear in 
{\it radiative\/} muon capture by a proton, for which interesting results
have been obtained in a recent TRIUMF experiment \cite{TRIUMF}.
A study of radiative muon capture in heavy baryon CHPT is in 
progress \cite{next}.

The external nucleon fields in our calculation are renormalized by defining a
wave function renormalization factor, $Z_N$, which is the residue of the
full nucleon propagator at the pole.  The derivation of $Z_N$
will be discussed in some detail, since our result differs from that of other
groups because of the different Lagrangian.  We will show, however, that within
our formalism our form of $Z_N$ is the one that
is necessary to ensure that the vector coupling is not renormalized, a result
required by CVC (Conserved Vector Current) theory, which of course ultimately
follows from QCD.

The muon capture reaction refers to a muon and proton with negligible relative
momentum, interacting to produce a neutron and neutrino,
\begin{equation}
   \mu + p \rightarrow \nu + n.
\end{equation}
The 4-momentum transfer in this process, $q = p_n - p_p$, satisfies
\begin{equation}
   q^2 = q_*^2 \equiv \frac{-m_\mu(m_p^2-m_n^2+m_\mu m_p)}{m_p+m_\mu} < 0.
\end{equation}
For nonradiative muon capture, $q_*^2 = -0.88 m_\mu^2$ and so is a small
parameter in the context of the CHPT expansion.

The general amplitude for muon capture can be parameterized as follows,
\begin{eqnarray}\label{ampl1}
   {\cal M} &=& \frac{-iG_\beta}{\sqrt{2}}\overline{u}({\bf p}_\nu)\gamma_
                  \alpha(1-\gamma_5)u({\bf p}_\mu)~\times \nonumber \\
              & & \overline{u}({\bf p}_n)\left[G_V(q^2)\gamma^\alpha
                  +\frac{iG_M(q^2)}{2m_N}\sigma^{\alpha\beta}q_\beta
                  -G_A(q^2)\gamma^\alpha\gamma_5
                  -\frac{G_P(q^2)}{m_\mu}q^\alpha\gamma_5\right]u({\bf p}_p),
\end{eqnarray}
where $G_V(q^2)$, $G_M(q^2)$, $G_A(q^2)$ and $G_P(q^2)$ are the form factors
to be studied, and
\begin{equation}
   \frac{G_\beta}{\sqrt{2}} = \frac{G_F{\rm cos}\theta}{\sqrt{2}}
                            = \frac{g_W^2{\rm cos}\theta}{8m_W^2}
                            = (0.8030 \pm 0.0008) \times 10^{-5}~{\rm GeV}^{-2}
                            ~.~~({\rm Ref.}~\cite{PDG})
\end{equation}
Here $m_N$ denotes the physical nucleon mass, $G_F$ is the Fermi constant, 
$\theta$ is the Cabibbo angle, and $m_W, g_W$ are the mass and weak
coupling constant of the W boson.  (The
proton-neutron mass difference is small and will be neglected.)
The so-called ``second class currents'' have not been shown, and do not arise 
in our CHPT
calculation.  

With the sign conventions of Eq.~(\ref{ampl1}), $G_A(q_*^2)$ and $G_P(q_*^2)$ 
are positive, matching the conventional positive sign for the parameter 
``$g_A$'' in 
heavy baryon CHPT.  The opposite signs for $G_A(q^2)$ and $G_P(q^2)$ have
almost always been used in non-CHPT studies of radiative and nonradiative 
muon capture, (e.g.\ Ref.~\cite{HWF}) and are used in Ref.~\cite{PDG}.

Following closely the notation of Ref.~\cite{EM96}, the heavy baryon chiral
Lagrangian is written in the form,
\begin{equation}
   {\cal L}^{EckM}_{\pi N} = \widehat{\cal L}^{(1)}_{\pi N} 
             + \widehat{\cal L}^{(2)}_{\pi N} + \widehat{\cal L}^{(3)}_{\pi N},
\end{equation}
where the superscripts on the right-hand side denote powers in the momentum
expansion.  The explicit forms of these three terms are given in Eqs.~(13),
(20) and (23) of Ref.~\cite{EM96}.  $\widehat{\cal L}^{(1)}_{\pi N}$ contains
two parameters, $F_0$ and $g_A$, which correspond to the chiral limits of the
pion decay and axial-vector coupling constants, respectively.  $\widehat{\cal
L}^{(2)}_{\pi N}$ introduces seven new parameters, labeled $a_1, a_2 \ldots,
a_7$, two of which will appear in our calculation, plus the nucleon mass in the
chiral limit.  Another 24 parameters, $b_1, b_2, \ldots, b_{24}$, arise from
$\widehat{\cal L}^{(3)}_{\pi N}$, and four of these will be present in our
study.

The pion field of Ref.~\cite{EM96} will be expressed here in an exponential 
representation,
\begin{equation}
   U = u^2 = {\rm exp}\left[\frac{i\sigma^a\pi^a}{F_0}\right]~~~,
       ~~{\rm Tr}[\sigma^a\sigma^b] = 2\delta^{ab}~.
\end{equation}
To match our ${\cal O}(p^3)$ calculation with the heavy baryon Lagrangian,
we will require terms through ${\cal O}(p^4)$ from the pure-meson chiral 
Lagrangian,
and for these we use the conventions of Gasser, Sainio and \v{S}varc\cite{GSS},
in particular the terms involving $l_3, l_4$ given in their Eq.~(5.9).
In both the meson and baryon sectors of the Lagrangian, 
the charged weak gauge bosons are included as external fields in the 
following manner,
\begin{equation}
   \ell_\mu = \frac{-g_W{\rm cos}\theta}{\sqrt{2}}
              \left(\begin{tabular}{cc} 0 & $W^+_\mu$ \\ $W^-_\mu$ & 0 \\
                    \end{tabular}\right)
              ~~~~,~~r_\mu = 0~.
\end{equation}
Throughout our calculation, we write the proton and neutron 4-momenta as
\begin{equation}\label{defvk}
   p_p^\alpha = m_{0N}v^\alpha+k_p^\alpha~~~,~~~
   p_n^\alpha = m_{0N}v^\alpha+k_p^\alpha+q^\alpha,
\end{equation}
where $m_{0N}$ is the bare nucleon mass, i.e.\ the nucleon mass in the
chiral limit, and $\alpha$ is the Lorentz index.  Any 
choices for $v^\alpha$ and $k_p^\alpha$ which satisfy the on-shell conditions
for the nucleons are valid, and the verification of this reparameterization
invariance \cite{LM} offers a check on our results.

In heavy baryon CHPT, the muon capture amplitude takes the form
\begin{equation}\label{ampl2}
   {\cal M} = \overline{u}({\bf p}_\nu)\frac{-ig_W}{2\sqrt{2}}\gamma_\alpha(1
             -\gamma_5)u({\bf p}_\mu)\frac{i}{m_W^2}\overline{n}_v({\bf p}_n)
                X^\alpha(q)n_v({\bf p}_p)~.
\end{equation}
We choose to denote the general field operator appearing in the relativistic
Lagrangian by $\psi(x)$ and for the matrix element use the Dirac 4-spinors 
$u({\bf p})$, with ${\bf p}$ being the
3-momentum, which arise in the expansion of the free field.  Similarly, 
$N_v(x)$ is the heavy 
baryon field and $n_v({\bf p})$ are the heavy baryon 4-spinors.  
$N_v(x)$ is obtained by making the usual heavy baryon transformation on
the Dirac field operator, i.e.,
\begin{equation}\label{HBtrans}
   N_v(x) = {\rm exp}[im_{0N}v{\cdot}x]\frac{1}{2}(1+v\!\!\!/)\psi(x).
\end{equation}
Some care must be exercised with the normalization. In matrix elements we 
always use normalized spinors, $\overline{u} u = \overline{n}_v n_v = 1$.

The function $X^\alpha(q)$ in Eq.~(\ref{ampl2}) has a one-particle-irreducible 
term and a pion pole term,
\begin{equation}\label{defineX}
   X^\alpha(q) = \Gamma^{(r)\,\alpha}_{pWn}(q)
            + \Gamma^{(r)}_{p{\pi}n}(q)\left[\frac{i}{q^2-m_\pi^2}\right]
              \Gamma^{(r)\,\alpha}_{W\pi}(q),
\end{equation}
where $m_\pi$ represents the physical pion mass, and the superscript ``$(r)$''
on each of the components means it has been renormalized individually and is 
finite.
To the order considered the renormalized pion propagator has the form of 
the free propagator, but with the physical mass.

We will now calculate the renormalized components of $X^\alpha(q)$, 
beginning with those from the pure-meson Lagrangian.
Working to ${\cal O}(p^4)$, the charged-pion, one-particle-irreducible,
unrenormalized, amputated, two-point, Green function 
is the sum of tree-level and one-loop contributions, which we compute using 
dimensional regularization in $d$ dimensions. 
\begin{equation}
   \Gamma_{\pi\pi}(q^2) = i(q^2-m_{0\pi}^2)
           +\frac{im_{0\pi}^2}{F_0^2}\left[2q^2l_4-2m_{0\pi}^2(l_3+l_4)
                           +\frac{(m_{0\pi}^2-4q^2)}{6(4\pi)^2}\left(R+
         {\rm ln}\left(\frac{m_{0\pi}^2}{\mu^2}\right)\right)\right].
\end{equation}
Here $m_{0\pi}$ is the bare pion mass and
$R$ contains the divergent piece of the loop graphs,
\begin{equation}
   R \equiv \frac{2}{d-4}-1+\gamma-{\rm ln}(4\pi)+{\cal O}(4-d).
\end{equation}
$\Gamma_{\pi\pi}(q^2)$ is related to the pion self energy $\Sigma(q^2)$ via
$\Gamma_{\pi\pi}(q^2)  = i(q^2-m_{0\pi}^2-\Sigma(q^2))$. Now to obtain the 
mass and wave function renormalization and the renormalized propagator 
we follow standard field theory techniques as described for example in 
Cheng and Li \cite{CL}. Thus we write for the full propagator, to the order to 
which we are working,
\begin{equation}\label{fpiprop1}
-\frac{1}{\Gamma_{\pi\pi}(q^2)} = \frac{i}{q^2-m_{0\pi}^2-\Sigma(q^2)}
=\frac{i}{q^2-m_{0\pi}^2-\Sigma(m^2_\pi)-(q^2-m_\pi^2)\Sigma^\prime (m^2_\pi)
-\tilde{\Sigma}(q^2)}~.
\end{equation}
To get the last equation we have expanded the self energy about the point 
$q^2=m_\pi^2$ so that $\Sigma(m^2_\pi)$ and $\Sigma^\prime (m^2_\pi)$ are 
respectively the 
value and derivative of $\Sigma(q^2)$ at that point and $\tilde{\Sigma}$ is 
zero in this particular case, but in general is the residual part which goes 
to zero at $q^2=m_\pi^2$ at least as fast as $(q^2-m_\pi^2)^2$.

The physical pion mass is obtained from the condition that the propagator have 
a pole at the physical mass, i.e. that
$\Gamma_{\pi\pi}(m_\pi^2) = 0$, giving $m_\pi^2=m_{0\pi}^2+\Sigma(m^2_\pi)$ or
\begin{equation}\label{physpimass}
   m_\pi^2 = m_{0\pi}^2\left[1+\frac{2m_\pi^2}{F^2}\left(l_3^r(\mu)+
             \frac{1}{4(4\pi)^2}{\rm ln}\left(\frac{m_\pi^2}{\mu^2}\right)
             \right)\right],
\end{equation}
where the parameter $l_3$ has been renormalized to absorb the divergence using
\begin{equation}
   l_3^r(\mu) = l_3+\frac{R}{4(4\pi)^2},
\end{equation}
and where we have anticipated that the difference between $F_0$ and the 
renormalized value $F$ is of higher order and have expressed all quantities
in the large brackets of Eq.~(\ref{physpimass}) 
in terms of physical quantities.

The full propagator can now be written, to the order to which we are working, 
as 
\begin{equation}\label{fpiprop2}
-\frac{1}{\Gamma_{\pi\pi}(q^2)} = \frac{i}{(1-\Sigma^\prime (m^2_\pi))
(q^2-m_\pi^2 - \frac{\tilde{\Sigma}}{1-\Sigma'(m^2_\pi)})} 
\equiv \frac{i Z_\pi}{q^2-m_\pi^2 -Z_\pi\tilde{\Sigma}}~.                 
\end{equation}
The value of the multiplicative wave function renormalization constant $Z_\pi$ 
is thus the residue of the propagator at the pole (i.e. at
the on-shell point) and is given by
\begin{equation}\label{defZpi}
   Z_\pi=\frac{1}{1-\Sigma^\prime (m^2_\pi)} 
   =\left[\frac{i(q^2-m_\pi^2)}{\Gamma_{\pi\pi}(q^2)}\right]_{q^2=m_\pi^2}
      = 1 - \frac{2m_\pi^2}{F^2}\left[l_4^r(\mu)+\frac{2}{3(4\pi)^2}R
      -\frac{1}{3(4\pi)^2}{\rm ln}\left(\frac{m_\pi^2}{\mu^2}\right)\right]~.
\end{equation}
The parameter $l_4$ has been renormalized using
\begin{equation}\label{renl4}
   l_4^r(\mu) = l_4-\frac{R}{(4\pi)^2}
\end{equation}
which we will justify below. Note that with this choice $Z_\pi$ is itself 
not finite, which is acceptable since, in contrast to  $m^2_\pi$, $Z_\pi$ 
is not an observable.

If we define a renormalized pion field by
\begin{equation}
   \pi^{(r)}(x) = \frac{\pi(x)}{\sqrt{Z_\pi}}~,
\end{equation}
then the two-point function for this renormalized field is finite, 
\begin{equation}
   \Gamma^{(r)}_{\pi\pi}(q^2) = Z_\pi\Gamma_{\pi\pi}(q^2) = i(q^2-m_\pi^2
     -Z_\pi\tilde{\Sigma})~,
\end{equation}
and the propagator for the renormalized field is the negative inverse
\begin{equation}
  - \frac{1}{\Gamma^{(r)}_{\pi\pi}(q^2)} = \frac{i}{q^2-m_\pi^2-
Z_\pi\tilde{\Sigma}}~.
\end{equation}
   At ${\cal O}(p^4)$ $\tilde{\Sigma}=0$, so that we obtain the 
renormalized propagator which  appears in Eq.~(\ref{defineX}).

The interaction between an unrenormalized pion field and a $W$-boson is
obtained from tree-level and one-loop diagrams using the meson chiral 
Lagrangian, and is
\begin{equation}
   \Gamma_{W\pi}^\alpha(q) = \frac{F_0}{2}{q^\alpha}g_W{\rm cos}\theta
      \left[1+\frac{2m_{0\pi}^2}{F_0^2}l_4-\frac{4m_{0\pi}^2}{3(4\pi{F_0})^2}
      \left(R+{\rm ln}\left(\frac{m_{0\pi}^2}{\mu^2}\right)\right)\right]~.
\end{equation}
The momentum $q$ flows from the $W$-boson to the pion, which is our convention 
for muon capture. To get the interaction of a renormalized pion, 
we multiply by $\sqrt{Z_\pi}$ to obtain 
\begin{equation}
   \Gamma^{(r)\,\alpha}_{W\pi}(q) = \frac{F}{2}{q^\alpha}g_W{\rm cos}\theta,
\end{equation}
where
\begin{equation}
   F = F_0\left[1+\frac{m_\pi^2}{F^2}\left(l_4^r(\mu)-\frac{1}{(4\pi)^2}
                  {\rm ln}\left(\frac{m_\pi^2}{\mu^2}\right)\right)\right]~.
\end{equation}
The renormalization of $l_4$ used in Eq.~(\ref{renl4}) above is required, 
since $F$  must be finite. 
Our normalization is such that $F = 92.4 \pm 0.3$ MeV\cite{PDG}.

Turning now to the nucleon Lagrangian, we use the nucleon two-point function to
determine the wave function renormalization factor and
the physical nucleon mass in a fashion analogous to that used for the 
pion.  Up to ${\cal O}(p^3)$ we find, for a nucleon with
four-momentum $p_p = m_{0N}v+k_p$,
\begin{equation}
   \Gamma_{NN}(p_p) = i\left[v{\cdot}p_p-m_{0N}-\Sigma(p_p)\right]
                    =i\left[v{\cdot}k_p-\Sigma(p_p)\right],
\end{equation}
where the tiny effects of isospin splitting (proportional to $m_d-m_u$)
have been neglected, and
\begin{eqnarray}\label{sigmaN}
   \Sigma(p_p) &=& -\frac{k_p^2}{2m_{0N}} -\frac{4a_3m_{0\pi}^2}{m_{0N}}
         - \frac{3g_A^2}{(4\pi{F_0})^2}\left[\frac{3}{4}
           v{\cdot}k_p\left(m_{0\pi}^2-\frac{2}{3}(v{\cdot}k_p)^2\right)
           \left(R+{\rm ln}\left(\frac{m_{0\pi}^2}{\mu^2}\right)-
           \frac{2}{3}\right)\right. \nonumber \\
    && ~~~~~~~~\left. +\frac{(v{\cdot}k_p)^3}{6}+(m_{0\pi}^2-
            (v{\cdot}k_p)^2)^{3/2}\left(
           \frac{\pi}{2}+{\rm arcsin}\left(\frac{v{\cdot}k_p}{m_{0\pi}}\right)
           \right)\right]
\end{eqnarray}
for $(v\cdot k)^2<m^2_\pi$.
In this expression the term involving $a_3$ is a contact term coming from a 
piece of $\widehat{\cal L}^{(2)}_{\pi N}$, the term with $1/(4\pi F_0)^2$ 
comes from loop contributions and the $k_p^2$ term comes from the term in 
$\widehat{\cal L}^{(2)}_{\pi N}$ proportional to $-\partial^2/2m_{0N}$ which 
involves no pions. We choose to include this as part of the interaction rather 
than part of the free Lagrangian and reserve for the free Lagrangian 
the $i v \cdot \partial$ term from $\widehat{\cal L}^{(1)}_{\pi N}$.

For the determination of the physical nucleon mass $m_N$ and renormalization
constant $Z_N$ we need the mass-shell condition, $p_p^2=m_N^2$, and $v \cdot
p_p =m_N$, where this second condition is motivated by the form of the lowest
order propagator, $i(v \cdot p_p -m_{0N})^{-1}$. Both conditions taken together
are equivalent to $p_p = m_N v$ which is what is usually stated in the 
literature.

$\Sigma(p_p)$ is a function of the four momentum $p_p$ and thus implicitly of $v$
and $k_p$. This dependence can be written in terms of the scalar variables 
$v\cdot p_p-m_N$ and $(p_p-m_Nv)^2$. In the vicinity of the pole at 
$p_p=m_N v$ these two variables are respectively first and second order in the
(small) distance from the pole.
 
We now proceed in exactly the same fashion as done for the pion and obtain 
formulas for the full nucleon propagator analogous to Eqs.~(\ref{fpiprop1}) 
and (\ref{fpiprop2}), i.e.,
\begin{eqnarray}
-\frac{1}{\Gamma_{NN}(p_p)} &=& \frac{i}{v \cdot p_p -m_{0N}-\Sigma(p_p)}
=\frac{i}{v \cdot p_p -m_{0N}-\Sigma(m_Nv)-(v \cdot p_p -m_N)\Sigma^\prime 
(m_Nv)-\tilde{\Sigma}} \nonumber \\ 
&=& \frac{i}{(1-\Sigma^\prime (m_Nv))
(v \cdot p_p -m_N - \frac{\tilde{\Sigma}}{(1-\Sigma^\prime (m_Nv))})} 
= \frac{i Z_N}{v \cdot p_p -m_N - Z_N 
\tilde{\Sigma}}~.
\end{eqnarray}
In these equations $\Sigma(m_Nv)$ and $\Sigma^\prime (m_Nv)$ are 
$\Sigma(p_p)$ and its derivative with respect to \mbox{$(v\cdot p_p-m_N)$} 
evaluated at $p_p=m_Nv$, i.e. at $(v \cdot p_p - m_N) = 0$,
and $\tilde{\Sigma}$ is the residual which goes to zero at the 
pole at least as fast as $(v \cdot p_p - m_N)^2$.

The evaluation of $\Sigma(m_Nv)$, $\Sigma^\prime (m_Nv)$, and $\tilde{\Sigma}$ 
requires an expansion of Eq.~(\ref{sigmaN}) but is 
relatively straightforward. To the order we are working the $a_3$ term 
contributes only to $\Sigma(m_Nv)$ whereas the loop piece contributes to all 
three. Note in particular that $\tilde{\Sigma}$ is not zero in this case. 
The $k_p^2$ term requires some discussion however. We can use $p_p-m_N v 
=k_p -(m_N-m_{0N})v$ to write 
\begin{equation}\label{kpsq}
\frac{k_p^2}{2m_{0N}}=\frac{(m_N-m_{0N})^2}{2m_{0N}}+\frac{(m_N-m_{0N})}
{m_{0N}}(v \cdot p_p-m_N) + \frac{(p_p-m_Nv)^2}{2m_{0N}}~.
\end{equation}
The first term on the right hand side contributes to $\Sigma(m_Nv)$ but 
is ${\cal O}(1/m_N^3)$, since as we shall see $(m_N-m_{0N})$ is 
${\cal O}(1/m_N)$, and so can be neglected. The second 
term however is only ${\cal O}(1/m_N^2)$ and will contribute to $\Sigma^\prime 
(m_Nv)$. Finally the third term contributes only to $\tilde{\Sigma}$.

One can now continue as before.  
The physical mass, $m_N$, is obtained from the requirement that the 
propagator have a pole at $p_p=m_N v$, i.e. at 
$m_N=m_{0N}+\Sigma(m_Nv)$, and we
find
\begin{equation}\label{nucmass}
   m_N = m_{0N}\left[1-\frac{4a_3m_\pi^2}{m_N^2}-\frac{3\pi{g_A^2}m_\pi^3}
                {2m_N(4\pi{F})^2}\right]~.
\end{equation}

The value of the multiplicative renormalization constant $Z_N$ is obtained as 
the residue of the propagator at the pole and is
\begin{equation}\label{defineZN}
   Z_N = \frac{1}{(1-\Sigma^\prime (m_Nv))}= \left[\frac{i(v \cdot p_p -m_N)}
{\Gamma_{NN}(p_p)}\right]_{p_p=m_N v}
       =  1+\frac{4a_3m_\pi^2}{m_N^2}-\frac{9g_A^2m_\pi^2}{4(4\pi{F})^2}
       \left(R+{\rm ln}\left(\frac{m_\pi^2}{\mu^2}\right)+\frac{2}{3}\right)~.
\end{equation}
As in the pion case, we can now define renormalized nucleon fields,
\begin{equation}
   N_v^{(r)}(x) = \frac{N_v(x)}{\sqrt{Z_N}},
\end{equation}
and the two-point function for these renormalized fields is finite 
($Z_N$ itself is not finite).
\begin{equation}
   \Gamma^{(r)}_{NN}(p_p) = Z_N\Gamma_{NN}(p_p).
\end{equation}
The propagator of the renormalized nucleon field is then
\begin{equation}\label{nucprop}
-\frac{1}{\Gamma^{(r)}_{NN}(p_p)} = \frac{i}{v \cdot p_p -m_N -
Z_N\tilde{\Sigma}} = \frac{i}{v \cdot k_p - (m_N-m_{0N}) -
Z_N\tilde{\Sigma}}~.
\end{equation}
For our calculation the propagator is only needed within loop diagrams,
and thus only the leading term in the propagator, $i/v{\cdot}k_p$, is needed.
However, the full expression, given in Eq.~(\ref{nucprop}), 
will be required for the nucleons in tree-level diagrams in more complicated 
processes.

The expression for the renormalization factor of the nucleon commonly found 
in the literature \cite{BKM95,PMR},
\begin{equation}
   Z_N^{\rm lit.} \equiv 1+\left[\frac{{\rm d}\Sigma^{\rm lit.}(v{\cdot}k_p)}
                            {{\rm d}v{\cdot}k_p}\right]_{v{\cdot}k_p=0}
                   = Z_N - \frac{4a_3m_\pi^2}{m_N^2} 
                     + {\cal O}\left(\frac{1}{m_N^3}\right)~,
\end{equation}
differs somewhat from our result.  
The extra term we obtain originates in the $(m_N-m_{0N})$ term from 
Eq.~(\ref{kpsq}). It involves $a_3$ by virtue of 
Eq.~(\ref{nucmass}).

The two definitions are not equivalent to the order of our calculation, and 
in our formalism 
it is $Z_N$, not $Z_N^{\rm lit.}$, which represents the full multiplicative 
renormalization function for the propagator.  Therefore $\sqrt{Z_N}$ is the
factor which exactly accounts for the renormalization of external nucleons in
the calculation of a matrix element and allows us to express all matrix 
elements in terms of fully renormalized and finite, one-particle-irreducible 
vertex functions.  

One way to see this difference explicitly
is to compute $G_V(q^2)$.  Because the vector current is conserved, the
calculation must give $G_V(0)=1$ to all orders in the $1/m_N$ expansion.
We will see that the use of $\sqrt{Z_N}$ for external nucleons 
in our formalism does satisfy
this constraint, whereas $\sqrt{Z_N^{\rm lit.}}$ does not.

The origin of this difference can be traced to a difference in the starting 
Lagrangian. The Lagrangian used in Refs.~\cite{BKM95,PMR} and in most previous
calculations contains a term which is transformed away using the equation of
motion in the form proposed by Ecker and Moj\v{z}i\v{s} \cite{EM96}. This term 
generates an $a_3$ term in $Z_N$ which just cancels the one we find. Of course
the physical results must nevertheless be the same, and in fact one finds 
that this term in
the Lagrangian also generates an additional $a_3$ term in the matrix element,
so that the final result for, say, $G_V$ is the same.

Although $Z_N$ is in itself not measurable, it does affect measurable
quantities. Thus one must always work within a consistent scheme in which 
the matrix elements, $Z_N$, and all other quantities are calculated
consistently from the same Lagrangian.

Clearly there is also some additional freedom in the way the finite parts of 
$Z_N$ are 
handled, even within a consistent scheme. In principle for example one could 
even within our formalism define a hybrid renormalization scheme in which 
$Z_N^{\rm lit.}$ accounts for all of the interactions on the external lines 
except the one involving $a_3$. Then in amplitudes one would have to add  
the one particle reducible diagrams involving the $a_3$ interaction on 
external legs. 
Alternatively, all insertions on external legs could be calculated explicitly.
While not as elegant or easy to do as the standard scheme used here, either of
these approaches
presumedly would lead to the same final result. Obviously a value of $Z_N$ has 
meaning only when what it includes and the scheme in which it is to be 
used is precisely defined.

Two more components of $X^\alpha(q)$ in Eq.~(\ref{defineX}) remain to be
determined.
The pion-nucleon vertex, renormalized by including the multiplicative factor
$Z_N\sqrt{Z_\pi}$, is
\begin{eqnarray}\label{PpiN}
   \Gamma^{(r)}_{p{\pi}n}(q) &=& \frac{-\sqrt{2}}{F}g_AS{\cdot}q\left(1+
         \frac{4a_3m_\pi^2}{m_N^2}\right)
   - \frac{\sqrt{2}g_A}{8m_N^2F}q{\cdot}(2k_p+q)\,S{\cdot}(2k_p+q) \nonumber \\
    && - \frac{4\sqrt{2}}{(4{\pi}F)^2F}S{\cdot}q\,m_\pi^2\left[b_{17}^r(\mu)
         -\frac{b_{19}}{2}-\frac{g_A^3}{4}-\frac{g_A}{4}(1+2g_A^2)\,\,
         {\rm ln}\left(\frac{m_\pi^2}{\mu^2}\right)\right],
\end{eqnarray}
and the $W$-nucleon vertex, including the factor $Z_N$, is
\begin{eqnarray}
   \Gamma^{(r)\,\alpha}_{pWn}(q) &=& \frac{-ig_W{\rm cos}\theta}{2\sqrt{2}}
                           \widetilde\Gamma_{pWn}^{(r)\,\alpha}(q), \\
   \label{PWNsub}
   \widetilde\Gamma^{(r)\,\alpha}_{pWn}(q) &=& (v^\alpha-2g_AS^\alpha)\left(1+
           \frac{4a_3m_\pi^2}{m_N^2}\right) \nonumber \\
     &&  + \frac{1}{2m_N}\left[(2k_p+q)^\alpha+2g_A
           S{\cdot}(2k_p+q)v^\alpha+8ia_6\epsilon^{\alpha\beta\gamma\delta}
           q_\beta v_\gamma S_\delta\right] \nonumber \\
     &&  + \frac{g_Aq^2}{4m_N^2}S^\alpha
         + \frac{4}{m_N^2}\left(a_6-\frac{1}{8}
           \right)i\epsilon^{\alpha\beta\gamma\delta}q_\beta {k_p}_\gamma 
           S_\delta
         + \frac{g_A}{2m_N^2}\left[S{\cdot}(k_p+q)
           k_p^\alpha+S{\cdot}k_p(k_p+q)^\alpha\right] \nonumber \\
     &&  - \frac{g_A}{4m_N^2}
           i\epsilon^{\alpha\beta\gamma\delta}q_\beta v_\gamma {k_p}_\delta
         - \frac{2q^2v^\alpha}{(4{\pi}F)^2}\left[
           b_7^r(\mu)+\frac{1}{12}(1+5g_A^2)\,\,{\rm ln}\left(
           \frac{m_\pi^2}{\mu^2}\right)\right] \nonumber \\
     &&  - \frac{8m_\pi^2}{(4{\pi}F)^2}S^\alpha\left[
           b_{17}^r(\mu)-\frac{g_A}{4}(1+2g_A^2){\rm ln}
           \left(\frac{m_\pi^2}{\mu^2}\right)\right] \nonumber \\
     &&  - \frac{2b_{23}}{(4{\pi}F)^2}(S{\cdot}q\,q^\alpha-q^2S^\alpha)
         - \frac{q^2v^\alpha}{18(4{\pi}F)^2}(1+17g_A^2) 
         + \frac{2g_A^3m_\pi^2}{(4{\pi}F)^2}S^\alpha \nonumber \\
     &&  + \frac{2v^\alpha}{(4{\pi}F)^2}\left[\frac{m_\pi^2}{3}(1+2g_A^2)
           -\frac{q^2}{12}(1+5g_A^2)\right]
           \int_0^1{\rm d}x\,{\rm ln}\left(1-x(1-x)\frac{q^2}{m_\pi^2}\right)
           \nonumber \\
     &&  -\frac{4{\pi}g_A^2m_\pi}{(4\pi{F})^2}
           i\epsilon^{\alpha\beta\gamma\delta}
           q_\beta{v}_\gamma{S}_\delta
           \int_0^1{\rm d}x\sqrt{1-x(1-x)\frac{q^2}{m_\pi^2}}~. 
\end{eqnarray}
We have used the standard definition,
\begin{equation}
   S^\alpha = \frac{i}{2}\gamma_5\sigma^{\alpha\beta}v_\beta ~.
\end{equation}
Notice that we have chosen to write Eqs.~(\ref{PpiN}) and (\ref{PWNsub}) in 
terms of the physical pion decay constant and the physical masses, rather 
than $F_0$ and the bare masses.
The remaining integrals in Eq.~(\ref{PWNsub}) can be done analytically for any
value of $q^2$, taking into account the appropriate 
boundary conditions in the propagators.  For example, at the muon capture 
point they become
\begin{eqnarray}
    \int_0^1{\rm d}x\,{\rm ln}\left(1-x(1-x)\frac{q_*^2}{m_\pi^2}\right) &=&
          -2+y(q_*^2)\,\,{\rm ln}\left(\frac{y(q_*^2)+1}{y(q_*^2)-1}\right), \\
   2 \int_0^1{\rm d}x\sqrt{1-x(1-x)\frac{q_*^2}{m_\pi^2}} &=&
          1+\sqrt{\frac{-q_*^2}{4m_\pi^2}}\,\,y^2(q_*^2)\,\,{\rm arccsc}\left(
          y(q_*^2)\right)
\end{eqnarray}
with
\begin{equation}
   y(q_*^2) \equiv \sqrt{1-\frac{4m_\pi^2}{q_*^2}}~.
\end{equation}
The effects of renormalization on all of the
parameters were studied by Ecker \cite{ECK94} using heat kernel techniques. The
form appropriate to the Lagrangian we use can be found in 
Ecker and Moj\v{z}i\v{s} \cite{EM96}, Table 1. We have used their results
\begin{eqnarray}
   b_7^r(\mu) &=& b_7+\frac{1}{12}(1+5g_A^2)R, \\
   b_{17}^r(\mu) &=& b_{17}-\frac{g_A}{4}(1+2g_A^2)R,
\end{eqnarray}
plus the fact that $b_{19}$ and $b_{23}$ do not get renormalized.  
This ensures that $\Gamma^{(r)}_{p{\pi}n}(q)$ and 
$\Gamma^{(r)\,\alpha}_{pWn}(q)$ are finite and do not depend on 
the renormalization scale $\mu$.

To obtain expressions for the muon capture form factors, we compare the two
forms for the amplitude, given in Eqs.~(\ref{ampl1}) and (\ref{ampl2}).
This comparison requires the well-known momentum dependence of relativistic 
4-spinors,
\begin{equation}
   u({\bf p}) = \left[\frac{m_N+p\!\!/}{\sqrt{2m_N(m_N+E)}}\right]u({\bf 0})~,
\end{equation}
where $E \equiv \sqrt{m_N^2+{\bf p}^2}$, and also their relation to the 
normalized heavy baryon 4-spinors,
\begin{equation}\label{litnlitu}
   n_v({\bf p}) = \sqrt{\frac{2m_N}{m_N+v \cdot p_p}}
\frac{(1+v\!\!\!/)}{2}u({\bf p}) =\left[1-\frac{k\!\!/_p}{2m_N}+\frac{(m_N-
m_{0N})}{2m_N}+\frac{k_p^2}{8m_N^2}+ {\cal O}(\frac{1}{m_N^3})\right] 
u({\bf p})
\end{equation}
as implied by Eq.~(\ref{HBtrans}) and the remark about 
normalization following it.
Eq.~(\ref{litnlitu}) is sufficient to rewrite Eq.~(\ref{ampl2}) in the form of
Eq.~(\ref{ampl1}). To reverse the procedure and rewrite Eq.~(\ref{ampl1}) 
in the form of Eq.~(\ref{ampl2}), note that for the specific choice of 
$v^\alpha$, $v^0=1$ and $v^i=0$ where $i=1,2,3$, we have 
$n_v({\bf p})=u({\bf 0})$. This allows us to write, for that choice 
of $v^\alpha$,
\begin{equation} \label{litulitn}
   u({\bf p}) = \left[\frac{m_N+p\!\!/}{\sqrt{2m_N(m_N+E)}}\right]n_v({\bf p})
    =\left[1+\frac{k\!\!/_p}{2m_N}-\frac{(m_N-m_{0N})}{2m_N}
+\frac{k_p^2}{8m_N^2}+ {\cal O}(\frac{1}{m_N^3})\right]n_v({\bf p}).
\end{equation}
The choice of $v^\alpha$ has no physical consequences, since the parameter
$k_p^\alpha$ will
adapt itself to the chosen $v^\alpha$ according to Eq.~(\ref{defvk}).
Also in some cases the algebra becomes simpler with the
choice of a particular Lorentz frame,
such as the lab frame where the muon and proton are stationary.

Whether or not such simplifying choices are made, the results are the same 
and we arrive at the following
expressions for the muon capture form factors.
\begin{eqnarray}
   G_V(q^2) &=& 1 - \left(a_6-\frac{1}{8}\right)\frac{q^2}{m_N^2}
         - \frac{q^2}{18(4{\pi}F)^2}(1+17g_A^2) \nonumber \\
     &&   - \frac{2q^2}{(4{\pi}F)^2}\left[
           b_7^r(\mu)+\frac{1}{12}(1+5g_A^2){\rm ln}\left(
           \frac{m_\pi^2}{\mu^2}\right)\right] \nonumber \\
     &&  + \frac{2}{(4{\pi}F)^2}\left[\frac{m_\pi^2}{3}(1+2
           g_A^2)-\frac{q^2}{12}(1+5g_A^2)
           \right]
           \int_0^1{\rm d}x\,{\rm ln}\left(1-x(1-x)\frac{q^2}{m_\pi^2}\right)
           \label{gV} \\
   G_M(q^2) &=& 4a_6 - 1 - \frac{4{\pi}g_A^2m_\pi{m}_N}{(4{\pi}F)^2}
           \int_0^1{\rm d}x\sqrt{1-x(1-x)\frac{q^2}{m_\pi^2}} \label{gM}, \\
   G_A(q^2) &=& g_A + \frac{4a_3g_Am_\pi^2}{m_N^2}
                - \frac{g^3_Am_\pi^2}{(4{\pi}F)^2} \nonumber \\
            &&  + \frac{4m_\pi^2}{(4{\pi}F)^2}\left[b_{17}^r(\mu)-\frac{g_A}{4}
                  (1+2g_A^2)\,{\rm ln}\left(\frac{m_\pi^2}{\mu^2}\right)\right]
                - \frac{b_{23}q^2}{(4{\pi}F)^2}, \label{gA} \\
   G_P(q^2) &=& \frac{2m_{\mu}m_N}{(m_\pi^2-q^2)}\left[G_A(q^2)
           - \frac{m_\pi^2}{(4{\pi}F)^2}(2b_{19}-b_{23})\right] \label{gP}.
\end{eqnarray}

To set the stage for our later calculation of radiative muon capture 
\cite{next}
it is necessary to evaluate parameters appearing in these expressions by 
comparison with other known experimental quantities. We do this in a fashion 
analogous to that of Ref.~\cite{BKKM}.
To extract numerical values for the parameters, we need only 
the first two terms in a $q^2$ expansion,
\begin{equation}
   G_X(q^2) \equiv G_X(0)\left[1+\frac{q^2}{6}\langle{r^2}\rangle_X
                   +{\cal O}(q^4)\right]~.
\end{equation}

It is clear from Eq.~(\ref{gV}) that $G_V(0) = 1$,
as required by conservation of the vector current.
The $a_3$ term in the first line of Eq.~(\ref{PWNsub}) which arose from $Z_N$
just cancels the similar term coming from the $(m_N-m_{0N})$ factor of
Eqs.~(\ref{litnlitu}) or (\ref{litulitn}) which are used in the extraction 
of $G_V(q^2)$.

Neglecting the electron mass, experimental data from neutron decay\cite{PDG} 
give $G_A(0) = 1.2601 \pm 0.0025$.

The value of $G_M(0)$ is related to the
nucleon magnetic moments, and implies a numerical value for $a_6$.
\begin{equation}\label{kapdiff}
   \kappa_p - \kappa_n = G_M(0) = 
   4a_6 - 1 - \frac{4\pi{m_\pi}m_NG_A^2(0)}{(4\pi{F})^2}
\end{equation}
where $\kappa_p = 1.7928$ and $\kappa_n = -1.9130$ are the anomalous
magnetic moments \cite{PDG}.
This gives
\begin{equation}\label{a6value}
   a_6 = 1.661 \pm 0.004,
\end{equation}
where the error is dominated by the
uncertainty in $F$ and where we have used the average nucleon mass and the 
charged pion mass in the evaluations.
This is in agreement with a recent precision determination of the 
${\cal L}^{(2)}_{\pi N}$ parameters \cite{BKM97}.
It should be noted that the error in Eq.~(\ref{a6value}) does not contain
any uncertainty relating to the convergence of the heavy baryon expansion.
For example, if we omit the last term in Eq.~(\ref{kapdiff}), corresponding
to the neglect of $\widehat{\cal L}^{(3)}_{\pi N}$, then the predicted value
of $a_6$ is reduced by 30\%.  The size of this reduction provides some
indication of the convergence of the chiral expansion, at least for this
particular quantity.

The muon capture form factors of Eqs.~(\ref{gV}-\ref{gP}) are related by CVC 
to the familiar electromagnetic form factors via
\begin{eqnarray}
   G_V(q^2) &=& F_1^p(q^2) - F_1^n(q^2)~, \\
   G_M(q^2) &=& \kappa_pF_2^p(q^2) - \kappa_nF_2^n(q^2)~.
\end{eqnarray}
Also, for $N=p,n$
\begin{eqnarray}
   {\cal G}^N_E(q^2) & \equiv & F^N_1(q^2) + 
                        \frac{q^2}{4m_N^2}\kappa_N F^N_2(q^2)~, \\
   {\cal G}^N_M(q^2) & \equiv & F^N_1(q^2) + \kappa_N F^N_2(q^2)~.
\end{eqnarray}
A simple parameterization which fits the electromagnetic data reasonably 
well\cite{Bosted} is the dipole,
\begin{equation}
   {\cal G}_E^n(q^2) \approx 0~~~,~~~
   {\cal G}_E^p(q^2) \approx \frac{{\cal G}_M^p(q^2)}{1+\kappa_p} \approx
                      \frac{{\cal G}_M^n(q^2)}{\kappa_n} \approx 
                      \left(1-\frac{q^2}{0.71{\rm GeV}^2}\right)^{-2}~.
\end{equation}
This dipole approximation gives
\begin{eqnarray}
   \langle{r^2}\rangle_V &=& 0.41~{\rm fm}^2~, \\
   \langle{r^2}\rangle_M &=& 0.72~{\rm fm}^2~.
\end{eqnarray}
The prediction of our CHPT calculation for $\langle{r^2}\rangle_M$ comes
from taking the $q^2$ expansion of $G_M(q^2)$, as given in Eq.~(\ref{gM}).
We obtain 
\begin{equation}\label{rMtheory}
   \langle{r^2}\rangle_M = \frac{2\pi{m}_N{G}_A^2(0)}{m_\pi(4\pi{F})^2G_M(0)}
                         = 0.523 \pm 0.004~{\rm fm}^2~.
\end{equation}
The CHPT prediction for $\langle{r^2}\rangle_M$ is about 30\% smaller than
the dipole estimate, which is not inconsistent with our expectation for
the effect of truncating the
chiral expansion at ${\cal O}(p^3)$, as discussed above for $a_6$ itself.
Using the dipole prediction for $\langle{r^2}\rangle_V$ as input to
Eq.~(\ref{gV}), we can obtain a value for the parameter $b_7^r$.
\begin{eqnarray}
   \langle{r^2}\rangle_V &=& \frac{3}{4m_N^2} - \frac{6a_6}{m_N^2}
       - \frac{1+7G_A^2(0)}{(4\pi{F})^2} - \frac{12}{(4\pi{F})^2}\left[
           b_7^r(\mu)+\frac{1}{12}(1+5G_A^2(0)){\rm ln}\left(
           \frac{m_\pi^2}{\mu^2}\right)\right] \\
   && \Rightarrow b_7^r(m_N) = -0.53 \pm 0.02.
\end{eqnarray}
Errors due to the chiral truncation or to the uncertainty in the dipole
prediction for $\langle{r^2}\rangle_V$ are not shown, though in fact they are
probably much larger than the other uncertainties included.

The quantity $\langle{r^2}\rangle_A$ has been measured in pion 
electroproduction\cite{Choi},        
\begin{equation}
   \langle{r^2}\rangle_A = 0.35 \pm 0.06~{\rm fm}^2
\end{equation}
and also in antineutrino-nucleon scattering\cite{Ahrens},
\begin{equation}
   \langle{r^2}\rangle_A = 0.42 \pm 0.04~{\rm fm}^2.
\end{equation}
Some corrections to the electroproduction analysis related to non-zero pion
mass actually bring that value into much closer agreement with the neutrino
scattering result \cite{BKM92}.
With Eq.~(\ref{gA}) and the latter value of $\langle{r^2}\rangle_A$, we obtain
an estimate of $b_{23}$,
\begin{equation}
   b_{23} = -\frac{1}{6}(4\pi{F})^2\langle{r^2}\rangle_AG_A(0) = -3.1 \pm 
0.3~.
\end{equation}

In principle, $b_{19}$ can now be determined from the experimental value of 
$\langle{r^2}\rangle_P$, which is, however, not very precise.
Instead, we observe that the pion-nucleon vertex function 
$\Gamma^{(r)}_{p{\pi}n}(q)$ is related to the renormalized pion nucleon
coupling constant via 
\begin{equation} 
-\sqrt{2}g_{\pi NN} \overline{u}({\bf p}_n)\gamma_5 u({\bf p}_p) =
\overline{n}_v({\bf p}_n)\Gamma^{(r)}_{p{\pi}n}(q)  n_v({\bf p}_p)
\end{equation}
and proceed in a fashion analogous to Ref.~\cite{BKM95}.
Applying the same evaluation to this as was used for the other couplings 
we find
\begin{equation}
g_{\pi NN} = \frac{m_N}{F}\left(G_A(0)-\frac{m_\pi^2 b_{19}}{8 \pi^2 F^2}
\right).
\end{equation}
Thus we see that $b_{19}$ is related to the so-called Goldberger-Treiman 
discrepancy
\begin{equation}
1-\frac{m_N G_A(0)}{F g_{\pi NN}}.
\end{equation}
Using $g_{\pi NN} = 13.0 \pm 0.1$ \cite{stoks}, corresponding 
to $q^2=m_\pi^2$, we find 
\begin{equation}
   b_{19} = -0.7 \pm 0.4~.
\end{equation}
We can now evaluate Eq.~(\ref{gP}) to obtain
\begin{equation}
   G_P(q_*^2) = 8.21 \pm 0.09
\end{equation}
which is in good agreement with the best value from non-radiative muon capture
\cite{Bardin},
\begin{equation}
   G_P(q_*^2) = 8.7 \pm 1.9~.
\end{equation}

In summary, we have computed the form factors of muon capture by a proton 
within the framework of the recently derived Ecker-Moj\v{z}i\v{s} 
${\cal O}(p^3)$ heavy baryon chiral Lagrangian, and extracted numerical
values for some of the Lagrangian's parameters from experimental data.  
The wave function renormalization factor for nucleons appropriate to this
Lagrangian and approach
was derived.

\acknowledgments

The authors would like to thank Thomas Hemmert for useful conversations.
S.S.\ would like to thank Norbert Kaiser for a useful discussion on
renormalization in HBCHPT.
This work was supported in part by the Deutsche Forschungsgemeinschaft,
the Natural Sciences and Engineering Research Council of Canada, and the 
NATO International Scientific Exchange Program.

\vfil\eject

\end{document}